\documentclass[conference]{IEEEtran}
\IEEEoverridecommandlockouts

\usepackage{graphicx}
\usepackage{hyperref}
\usepackage{float}
\usepackage{cite}
\usepackage{booktabs}
\usepackage{balance}
\usepackage{subfig}

\def\BibTeX{{\rm B\kern-.05em{\sc i\kern-.025em b}\kern-.08em
    T\kern-.1667em\lower.7ex\hbox{E}\kern-.125emX}}
\begin{document}
\IEEEpubid{\makebox[\columnwidth]{979-8-3315-6165-9/26/\$31.00 \textcopyright 2026 IEEE \hfill} \hspace{\columnsep}\makebox[\columnwidth]{}}

\title{Interactive Augmented Reality-enabled Outdoor Scene Visualization For Enhanced Real-time Disaster Response
\thanks{This work has received funding from the European Union’s Horizon Europe research and innovation programme under Grant Agreement No. 101168042 project TRIFFID (auTonomous Robotic aId For increasing First responders Efficiency). The views and opinions expressed in this paper are those of the authors only and do not necessarily reflect those of the European Union or the European Commission.}
}

\author{
  \IEEEauthorblockN{
    Dimitrios Apostolakis\IEEEauthorrefmark{1},
    Georgios Angelidis\IEEEauthorrefmark{1}\IEEEauthorrefmark{2},
    Vasileios Argyriou\IEEEauthorrefmark{3},\\
    Panagiotis Sarigiannidis\IEEEauthorrefmark{4},
    Georgios Th. Papadopoulos\IEEEauthorrefmark{1}\IEEEauthorrefmark{2}
  }

  \IEEEauthorblockA{
    \IEEEauthorrefmark{1}Department of Informatics and Telematics, Harokopio University of Athens, Athens, Greece
  }

  \IEEEauthorblockA{
    \IEEEauthorrefmark{2}Archimedes, Athena Research Center, Athens, Greece
  }

  \IEEEauthorblockA{
    \IEEEauthorrefmark{3}Department of Networks and Digital Media, Kingston University, London, United Kingdom
  }

  \IEEEauthorblockA{
    \IEEEauthorrefmark{4}Department of Electrical and Computer Engineering, University of Western Macedonia, Kozani, Greece
  }

  \IEEEauthorblockA{
    Emails: \{it2022004, gangelidis, g.th.papadopoulos\}@hua.gr, \\
    Vasileios.Argyriou@kingston.ac.uk, psarigiannidis@uowm.gr
  }
}

\maketitle
\IEEEpubidadjcol

\begin{abstract}
A user-centered AR interface for disaster response is presented in this work that uses 3D Gaussian Splatting (3DGS) to visualize detailed scene reconstructions, while maintaining situational awareness and keeping cognitive load low. The interface relies on a lightweight interaction approach, combining World-in-Miniature (WIM) navigation with semantic Points of Interest (POIs) that can be filtered as needed, and it is supported by an architecture designed to stream updates as reconstructions evolve. User feedback from a preliminary evaluation indicates that this design is easy to use and supports real-time coordination, with participants highlighting the value of interaction and POIs for fast decision-making in context. Thorough user-centric performance evaluation demonstrates strong usability of the developed interface and high acceptance ratios.
\end{abstract}


\section{Introduction}

Effective emergency response and Search and Rescue (SAR) missions often occur under extreme time pressure and in challenging conditions \cite{cani2025triffid}. In these contexts, First Responders (FRs) must maintain strong Situational Awareness (SA) \cite{Endsley00} to support both sound decision-making and operational safety \cite{Nasar23, Tveit19, Lutz18}. Traditionally, commanders and field teams depend on 2D maps, voice-based communication, and contradictory data feeds to construct a Common Operational Picture (COP) \cite{Agrawal21}. Yet, converting this two-dimensional information into an accurate understanding of the three-dimensional real-world places a heavy cognitive burden on users by forcing them to perform mental transformations and assumptions about geometry and spatial relationships. In time-critical scenarios, these added demands can slow situation assessment and contribute to avoidable navigation or coordination errors \cite{Vuckovic22, Sharma19}.

\begin{figure}[t]
    \centering
    \includegraphics[width=\columnwidth]{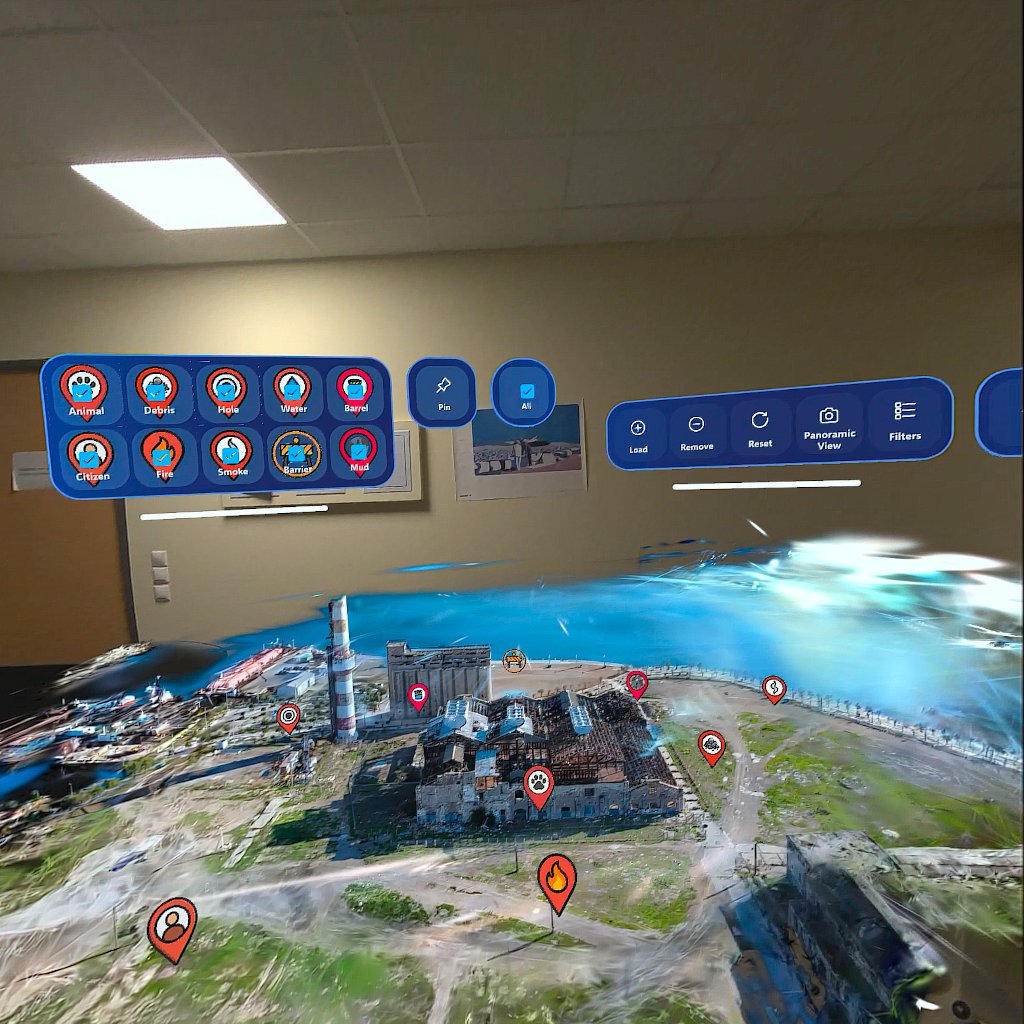}
    \caption{Overview of the AR-enabled disaster scene.}
    \label{fig:overview}
\end{figure}

To overcome these limitations, Augmented Reality (AR) has gained traction as an effective means to connect digital information with the physical environment \cite{Khanal22, Chmielewski19, Hancko25, Beroggi95}. By overlaying key data directly onto the user’s visual field, AR interfaces promote “recognition rather than recall,” in line with one of Nielsen’s core usability heuristics \cite{Nielsen94}. Nevertheless, AR is not merely a visualization comfort but a safety-critical interface, meaning that users may rely on it for time-sensitive judgments (e.g., navigation and hazard avoidance), so UI errors or latency can translate into operational mistakes and safety risks. Performance therefore matters beyond raw throughput \cite{Mirbabaie19}. Latency and unstable frame timing can cause spatial hints to lag or drift relative to head motion, undermining distance and alignment judgments and complicating safe coordination. Yet, depicting complex disaster environments requires rendering enormous amounts of geometric detail (e.g., rubble, fires, structural degradation), which mobile or wearable hardware handles inefficiently.

\begin{figure}[t]
    \centering
    \includegraphics[width=\columnwidth]{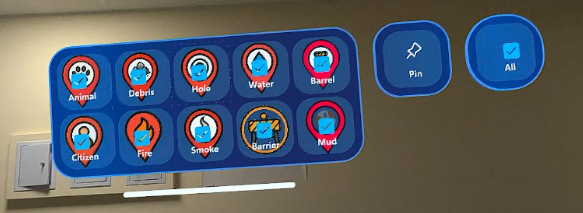}\\
    {\small (a) Filter menu}\\[8pt]
    \includegraphics[width=\columnwidth]{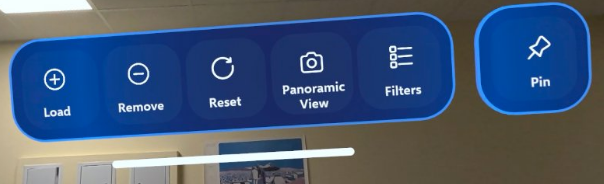}\\
    {\small Main menu}
    \caption{Detailed view of the functionalities.}
    \label{fig:components}
\end{figure}

Recently, 3DGS \cite{Kerbl23} has gained attention as an efficient scene-reconstruction technique, delivering photorealistic rendering with speeds that exceed those of traditional Neural Radiance Fields (NeRFs) \cite{Mildenhall21, Liu25, Blanchard23, Lee24}. Even though 3DGS is capable of delivering the visual accuracy required for achieving precise damage assessment, its efficiency as a credible AR solution for real-time applications is not yet carefully investigated. In particular, there is still limited interpretation regarding the scalability of 3DGS on resource-constrained devices. Moreover, disaster environments are naturally dynamic: hazards can emerge or spread, access routes may become blocked or rerouted, and damaged structures can further decay over time. Consequently, a static reconstruction may become outdated within minutes and risk delivering an inaccurate picture of the scene, underscoring the need for timely, real-time updates to maintain reliable situational awareness.

In this work, a user-centered AR-based Human-Computer Interface (HCI) is introduced, tailored to disaster response operations. This approach leverages 3DGS to achieve precision-level visualization of evolving scenes while following usability best practices to reduce cognitive load on FRs. The main contributions of this paper are as follows:

\begin{itemize}
    \item A system architecture that connects an AR client to a lightweight backend to support real-time dynamic updates of 3DGS scenes without interrupting user interaction.
    \item An AR interface design that prioritizes SA through not-intrusive visual elements, applicable to both PC-AR and standalone AR hardware.
    \item A comprehensive performance evaluation across mobile, standalone and PC-AR configurations, highlighting hardware constraints for large-scale disaster scenes, complemented by a pilot user study on usability and coordination support.
\end{itemize}

The rest of this paper is organized as follows: Section \ref{sec:related} reviews the related works on 3D scene reconstruction and the use of VR-based visualization in emergency response situations. Section \ref{sec:methodology} presents the methodology of the proposed system. Section \ref{sec:experiments} details the experimental setup and reports the performance together with findings from the pilot user study. Section \ref{sec:discussions} discusses the main limitations and drawbacks encountered, and outlines directions for future work. Finally, Section \ref{sec:conclusions} concludes the paper and summarizes the key findings.

\section{Related Work} 
\label{sec:related}

The potential of immersive technologies to aid emergency decision-making was identified early, with seminal studies showing their usefulness in rescue operations as early as the mid-1990s \cite{Beroggi95}. However, over the past decade, the field has experienced a paradigm shift. Propelled by progress in photogrammetry and computer vision \cite{konstantakos2025self, alimisis2025advances, cani2026illicit}, research has advanced from basic simulations to highly accurate Digital Twins and AI-supported SA \cite{Khanal22, Zhu21}.

Nevertheless, much of the literature that currently exists examines the use of immersive technologies from the perspective of training. Hancko et al. \cite{Hancko25} present a review that describes the use of immersive technologies to train firefighters, highlighting the possibilities that the technology has to offer, including the potential to improve safety, repeatability, and scalability of training, as well as the benefits that are applicable to the fire service, such as improved SA, decision-making, and teamwork, and the estimation of cost efficiency. Other literature, such as that presented by Lovreglio et al. \cite{Lovreglio21}, who compared the use of VR to conduct training on the use of fire extinguishers with video-based training, found that participants who used the VR reported better knowledge and confidence, while Gong et al. \cite{Gong15} developed an earthquake training system that places the participant in a virtual ecosystem, enabling the researcher to observe and evaluate the reaction to the situation.

In parallel, several studies show how integrating high-quality geospatial data with interactive 3D environments can support crisis management workflows. Tully et al. \cite{Tully15} introduced the Project Vision Support framework, which integrates LiDAR, OpenStreetMap, and Ordnance Survey data within a game engine to generate hybrid, high-resolution 3D maps for crisis response. Similarly, Velev et al. \cite{Velev19} discuss how UAVs paired with AR can empower near real-time disaster mapping, allowing responders to identify critical locations and assess structural damage from a safer distance. 

Generally, these works point to an assembled ecosystem where data-rich 3D reconstruction, immersive interfaces, and rapid mapping pipelines boost each other, supporting both operational decision-making during incidents and realistic, repeatable training for emergency crew. Taken together, these studies highlight the value of immersive 3D environments for situational awareness and decision support. However, most reported systems emphasize training scenarios or static/post-disaster reconstructions. The proposed system shifts the role of VR from training to operations. VR becomes a real-time workspace for coordination, where the 3D scene is updated dynamically as new information arrives.

\section{Methodology}
\label{sec:methodology}

A back-end service is used to reconstruct a 3D scene from UAV data as it is collected. The backend processes the incoming measurements and generates a `.ply' file that represents the reconstructed environment. This file is then streamed to the front-end via WebSockets, sanctioning near real-time updates. More details about the back-end development are described here \cite{Angelidis26}.

On the front-end side, the system uses an interactive engine designed for building immersive experiences. A lightweight user interface allows the user to interact with the scene (e.g., inspection and navigation in it). The reconstructed 3D environment is rendered directly from the streamed `.ply' data and can be visualized not only on a standard display, but also through VR headsets for an immersive, first-person view of the scene.

\subsection{AR Interface and Interaction Design}
\label{sec:uiux}

The proposed interaction framework is designed with the overall goal of optimizing the balance between SA and cognitive load in high-pressure environments. Grounded on Cognitive Load Theory \cite{Hollender10} and established HCI guidelines \cite{Nielsen94}, the system follows an adaptive, minimalist approach that emphasizes task-critical information, while limiting visual clutter and redundant interaction steps. Consistent with the “recognition rather than recall” heuristic, essential information is presented through in-view contextual cues and explicit highlighting of salient POIs, thereby reducing unnecessary mental effort. This design choice is supported by recent work showing that the “detail-in-context” visualization strategy significantly enhances the operator's ability to form an accurate mental representation of remote environments \cite{Bakzadeh25}. In addition, the system operates as a cognitive artifact to foster a shared mental model (common ground) among distributed teams. This functionality is vital for counteracting performance losses under stress in unfamiliar settings \cite{Migliorini21} and for supporting the interpretation of complex disaster scenarios through high-fidelity simulation \cite{Zhu21}.

\subsubsection{Scene Interaction}

Because disaster scenes often cover large areas, physical movement alone is insufficient for a fast inspection. Rather than depending exclusively on teleportation, a WIM \cite{Stoakley95} manipulation technique is integrated. Users manipulate the map directly via hand controllers. With these controllers, they can perform standard affine transformations:

\begin{itemize}
    \item Scaling: Scaling is achieved through bimanual controller interaction, allowing users to smoothly transition between a global and a local view of the environment. This enables fluid switching between a high-level overview of the site and a close-up examination of specific rubble.
    \item Rotation and Translation: Users can rotate and reposition the map to reveal otherwise occluded regions without having to physically move around the room, which is especially advantageous in tight operational settings.
\end{itemize}

Finally, the interface includes a reset function that allows users to restore all scene transformations (e.g., translation, rotation, and scale) to their default state, for quick recovery from unintended manipulations during time-critical operations.

\subsubsection{Visualization of POIs}

Disaster environments are intrinsically disordered. To support interpretation, the system introduces semantic layers. Through UI, users can toggle individual hazard classes (e.g., “Fire", “Smoke", “Debris", etc.) on or off. In doing so, commanders can focus on mission-critical information without being distracted by irrelevant visual elements. At the same time, this approach allows task-oriented data exploration, while maintaining a critical understanding of the overall situation \cite{Bakzadeh25}.
\subsubsection{Spatial awareness via Passthrough}
To keep operators safe and to make teamwork easier on site, the system uses mixed‑reality video passthrough so that users can still see what is happening around them while viewing digital overlays. Unlike fully immersive VR, passthrough helps users feel less cut off and makes it easier to talk and coordinate with nearby colleagues. This lets an operator quickly switch from reading digital information to taking real‑world action, an important capability in fast‑changing emergency situations.
For configurations that lack passthrough capabilities, limitations of which are discussed in Section \ref{sec:discussions}, the interface reverts to a VR environment. To balance the loss of physical context, the system provides locomotion controls \cite{Pausch95} through the controllers for detailed exploration, so the user could navigate “inside" the 3D scene and inspect the terrain with the same level of immersion as an on-site responder.

\section{Experiments and Evaluation}
\label{sec:experiments}

To verify the effectiveness of the system's interaction, navigation, and visualization capabilities, a preliminary evaluation was conducted involving 12 participants, a sample size of $10 \pm 2$, selected according to \cite{Hwang10}. This evaluation focused on assessing how well users could interact with the 3D reconstructed environment, navigate through the disaster scene, and perceive the visual quality of the reconstructed data. 

Also, a performance experiment was executed across mobile AR, PC-AR and standalone AR headsets, to check the rendering computational load and performance of a 3D post-disaster reconstructed scene under different hardware constraints. Frame rate (FPS) served as the main evaluation metric, since maintaining high and stable FPS is essential for user comfort and conscious interaction, especially in immersive VR/AR settings \cite{Geris24, Zielinski15}. Furthermore, frame time was measured, which is simply how long it takes to draw a single frame on screen. Lower and more consistent frame times make motion look smoother and the system feel more responsive.

\subsection{Experimental Setup}

\begin{figure}[H]
    \centering
    \includegraphics[width=1.0\linewidth]{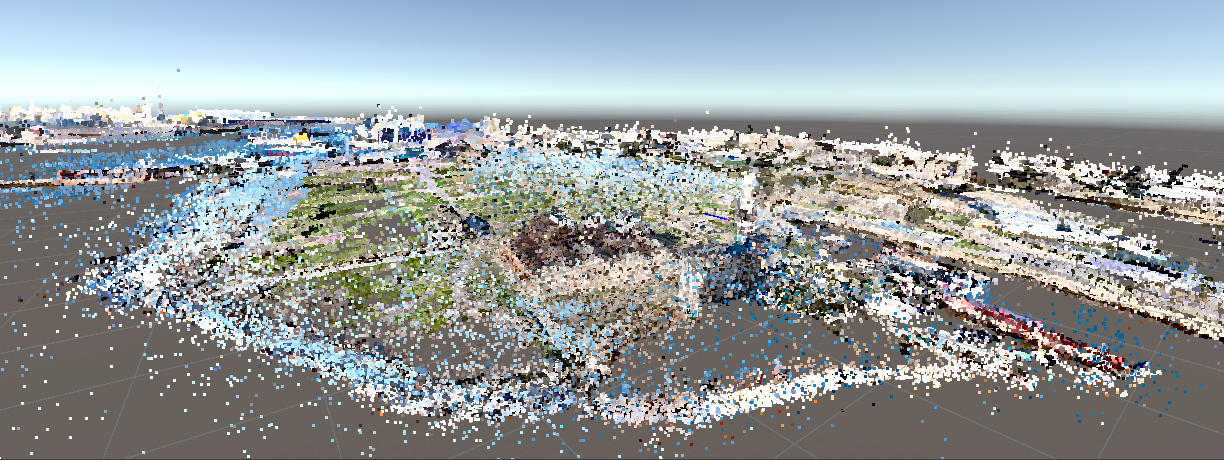} 
    \caption{Snapshot of the 1.144 million debug points of the 3DGS sample used for the evaluation study from  the Unity Editor}
    \label{fig:dataset}
\end{figure}

For our purpose, because the custom rendering pipeline for 3DGS was designed in Unity \cite{Kleinbeck25, Pranckevičius24}, Unity 2022.3 is used as the front-end, while UI/UX are reused components from MRTK3 \cite{Microsoft23}, as they align with the design guidelines discussed in Section \ref{sec:uiux}. For the benchmarking process, a high-fidelity 3D reconstruction based on Gaussian Splatting is tested. This scene consists of 1.144.277 Gaussian splats, being depicted in Fig \ref{fig:dataset}, capturing complex geometries, such as debris, concrete structures, and scattered hazards. This dataset was chosen to stress-test the sorting algorithms, as the scene visually resembles a disaster-like environment.

Three hardware setups are evaluated:
\begin{itemize}
    \item Mobile AR used a Samsung S24 FE (Exynos 2400e, Samsung Xclipse 940,
 8 GB RAM), capturing a lightweight, handheld AR scenario.
    \item Standalone VR used a Meta Quest 3 (Adreno 740 GPU), representing an untethered, fully immersive VR experience.
    \item PC-AR relied on a desktop workstation with an Intel Core i7 and an NVIDIA RTX 3080 Ti (12 GB VRAM), streaming to the Meta Quest 3 via Oculus Link, which reflects to the proposed base-station architecture.
\end{itemize}
Mobile and standalone  VR deployments rely on Vulkan, due to API support for compute operations. Desktop deployment utilizes DirectX 12 to maximize GPU throughput. 

\subsection{Performance Experiment Results}

Table \ref{tab:fps} summarizes the rendering performance and runtime characteristics across the three tested devices. In terms of frame rate, the PC-AR maintained 72 FPS, while both mobile and the standalone VR headset remained below 10 FPS ($\leq$7 and $\leq$9 FPS, respectively).

% FPS table
\begin{table}[H]
    \centering
    \caption{Devices and FPS comparisons}
    \label{tab:fps}
    \footnotesize
    \begin{tabular}{l l c}
        \toprule
        \textbf{Device} & \textbf{GPU} & \textbf{FPS (Avg)} \\
        \midrule
        Mobile & Samsung Xclipse 940 & 7 \\
        Meta (AR) & Adreno 740 & 9 \\
        Desktop & NVIDIA GeForce RTX 3080 Ti & \textbf{72} \\
        \bottomrule
    \end{tabular}
\end{table}

The memory and timing results in Table \ref{tab:memory} highlight clear differences in how each device uses resources. On the PC, the application shows the highest overall RAM footprint, with 1485 MB reserved and 937 MB allocated, while still achieving a relatively low frame time of 13.8 ms. In contrast, Mobile and standalone AR use less memory overall, with 387 MB reserved and 229 MB allocated on mobile and 504 MB reserved and 411 MB allocated on standalone, but they run with much higher frame times at 144.4 ms and 111.3 ms, respectively. Mono memory follows the same pattern, remaining much higher on PC at 527 MB and dropping to 8 MB on mobile and 5 MB on standalone.

% Memory & ms table
\begin{table}[H]
    \centering
    \caption{Memory Usage and Frame Time (ms) Across devices}
    \label{tab:memory}
    \begin{tabular}{l c c c c}
        \toprule
        \textbf{Device} & \textbf{Reserved} & \textbf{Allocated} & \textbf{Mono} & \textbf{Frame Time (ms)} \\
        \midrule
        Desktop & 1485 MB & 937 MB & 527 MB & \textbf{13.8} \\
        Mobile & 387 MB & 229 MB & 8 MB & 144.4 \\
        Meta (AR) & 504 MB & 411 MB & 5 MB & 111.3 \\
        \bottomrule
    \end{tabular}
\end{table}

\subsection{User Evaluation Results}

The user evaluation results in Table \ref{tab:user_results} show that the mean of user perception ease of map interaction was 4.42, menu accessibility was 4.36, and usefulness for coordination was 4.55, while the standard deviations was 0.82, 0.84, and 0.52, respectively.

\begin{table}[H]
    \centering
    \caption{User Evaluation Results (N=12) on a 5-points Likert scale}
    \label{tab:user_results}
    \begin{tabular}{l c c}
        \toprule
        \textbf{Evaluation Criterion} & \textbf{Mean Score ($\mu$)} & \textbf{SD ($\sigma$)} \\
        \midrule
        Ease of Map Interaction & 4.42 & 0.82 \\
        Menu Accessibility & 4.36 & 0.84 \\ 
        Usefulness for Coordination & 4.55 & 0.52 \\ 
        \bottomrule
    \end{tabular}
\end{table}

\section{Discussion, Limitations and Future work}
\label{sec:discussions}

The results of the user evaluation study showed that the user interface was highly usable, with positive ratings for usability and perceived support for coordination, as all the mean ratings were above 4 out of 5, and the highest agreement was obtained for the usefulness of the interface in coordination. Additionally, based on the performance experiments, the PC-AR configuration is currently the most suitable option for visualization, achieving a stable 72 FPS with a 13.8 ms frame time. In this setup, a desktop workstation at the ground station absorbs the computational cost required to render the 3DGS scene, while the view is streamed to the headset. However, according to Meta’s documentation, passthrough for applications connected to a PC via Oculus Link is intended for development use \cite{Meta25} and is available only when running through the Unity Editor, not in a standalone Windows build. Prior work \cite{Pranckevičius24, Kleinbeck25} reports stable 72 FPS up to approximately 400k splats, which indicates that a standalone headset deployment can also be feasible under appropriate scene complexity, while additionally supporting passthrough. Finally, because splats are rendered as an image-based representation rather than physical geometry, interaction is handled using a simplified approach: a collider is placed around the boundary of the map and manipulation is performed based on that proxy collider.

In future work, based on the received feedback, the system will be extended with map annotation capabilities, providing the coordinator the ability to mark and comment on areas of interest directly within the reconstructed scene. Multi-user support will be explored, allowing multiple coordinators to simultaneously view and collaborate in the same environment. Beyond headset-based interaction, interaction patterns that do not require wearable equipment will be investigated. Additionally, POIs will be made interactive so that coordinators can select them and view additional useful information. Moreover, the system will be evaluated with FRs to validate its effectiveness in simulated operational scenarios. Furthermore, the system will be connected with real-time AI-enabled event detectors \cite{linardakis2025survey, foteinos2025visual, linardakis2024distributed}, so as to facilitate scenarios involving Human-Robot Interaction \cite{papadopoulos2021towards, papadopoulos2022user, moutousi2025tornado} and broader security response incidents \cite{mademlis2024invisible}, while also supporting the necessary eXplainable Artificial Intelligence (XAI) pipelines \cite{rodis2024multimodal, evangelatos2025exploring}.

\section{Conclusions}
\label{sec:conclusions}

In this paper, a user-centered AR/VR interface for disaster management was presented, that uses 3DGS to visualize high-accuracy reconstructions while prioritizing SA and low cognitive load. By combining a lightweight interaction design (WIM-based navigation and semantic POIs toggles) with an architecture that supports streaming updates of reconstructed scenes, the system is positioned as an operational workspace rather than a training tool. This aims to help coordinators and field teams to build and to maintain a shared understanding of rapidly changing environments.

A preliminary evaluation with 12 participants provides early evidence that the interaction design is effective and easy to use. Participants rated 3D map interaction and menu access highly, reported that POIs and POI filtering support faster, more focused decision-making. They also perceived the system as useful for real-time coordination, suggesting that the interface can support shared understanding in time-critical scenarios.

The performance study highlights a key practical outcome: large scale 3DGS scenes remain challenging for mobile and standalone headsets when rendered locally. In contrast, for scenes that contain massive amounts of splats, PC-AR sustained a stable 72 FPS, while mobile AR and standalone AR stayed below 10 FPS. This gap indicates that, for high-detail disaster reconstruction, a base-station approach (rendering on a workstation, streaming to a headset) is currently the most reliable option for responsive interaction and comfortable frame timing. At the same time, the observed limitations, especially around passthrough availability in PC-AR configuration and the simplified interaction afforded by splat-based representations, show that system-level constraints are as important as raw reconstruction quality, when deploying AR for safety-critical use.

Future work will extend the system towards real operational deployment by supporting multi-user collaboration, in-scene annotations for coordination, and richer POI interaction to surface additional contextual information.

\balance

\bibliographystyle{IEEEtran}
\bibliography{references}
\end{document}